\documentclass[twocolumn]{aastex7}
\usepackage{CJK}
\usepackage{amsmath}
\usepackage{graphicx}

\usepackage[utf8]{inputenc}

\usepackage{amsmath}
\renewcommand{\thesubsection}{\Roman{subsection}}
\usepackage{epstopdf}
\usepackage{commath}
\usepackage{natbib}
\usepackage{color}
\usepackage{mathtools}
\usepackage[toc,page]{appendix}
\usepackage{bm}
\usepackage{hyperref}

  \renewcommand{\d}{\mathrm d}

\begin{document}

\title{Kinetic wave activity and proton heating in 3D hybrid simulations of decaying balanced and imbalanced Alfvénic turbulence}

\author[orcid=0000-0001-7063-2511]{C.A. González}
\email{carlos.gonzalez1@austin.utexas.edu}
\affiliation{Department of Physics, The University of Texas at Austin, Austin, TX 78712, USA}
\correspondingauthor{C.A. González}
\author[orcid=0000-0002-4625-3332]{T. A Bowen}
\email{tbowen@Berkeley.edu}
\affiliation{Space Sciences Laboratory, University of California,Berkeley, CA 94720-7450, USA}

\author[orcid=0000-0001-9202-1340]{A. Mallet}
\email{alfred.mallet@berkeley.edu}
\affiliation{Space Sciences Laboratory, University of California,Berkeley, CA 94720-7450, USA}

\author[orcid=0000-0003-2880-6084]{A. Tenerani}
\email{Anna.Tenerani@austin.utexas.edu}
\affiliation{Department of Physics, The University of Texas at Austin, Austin, TX 78712, USA}

\begin{abstract}
We investigate turbulent dynamics from large to sub-proton scales, the kinetic wave activity and resulting proton heating in three-dimensional hybrid-kinetic simulations of freely decaying balanced and imbalanced Alfvénic turbulence. We considered initial fluctuations with same amplitude and spectral shape but different degrees of correlation between velocity and magnetic field perturbations (i.e., normalized cross-helicity), which define the corresponding turbulent regimes. We found that our balanced turbulence simulation exhibits stronger energy dissipation and more efficient proton heating compared to the imbalanced case. On the large fluid scales, the two simulations undergo distinct nonlinear dynamic and energy cascades, leading to different spectral properties once turbulence is well developed. At sub-proton scales, fluctuations are predominantly oblique with right-hand polarization, a property common to both regimes. In contrast, only the imbalanced simulation displays substantial wave power parallel to the guide magnetic field, with clear signatures of ion-cyclotron waves. These waves are excited in regions with strong temperature anisotropy that arise during the initial nonlinear steepening of Alfvénic fluctuations, a process that simultaneously generates parallel-propagating proton beams. Furthermore, non-Maxwellian features emerge in velocity space, indicating different heating mechanisms in balanced and imbalanced regimes and providing evidence for the role of kinetic instabilities in regulating the turbulent dynamics. Overall, these results show that cross-helicity is a key parameter controlling the large-scale evolution, kinetic activity, and energy dissipation in collisionless turbulent plasmas.

\end{abstract}

\section{Introduction}
\label{intro}

Collisionless magnetized plasmas, ubiquitous in space and astrophysical environments, are usually in a turbulent state \citep{higdon1984density,quataert1998particle,elmegreen2004interstellar,schekochihin2007turbulence}. Understanding how energy is transported and dissipated is crucial for a complete and accurate modeling of these systems. Extensive observations of the solar wind — our most accessible weakly collisional plasma that is produced by the continuous outflow from the Sun’s corona — have revealed a highly complex dynamics in which turbulent fluctuations play an important role heating and accelerating the wind \citep{tu1995mhd,bruno2013solar}. However, a fully consistent picture of the physical mechanisms governing dissipation and damping of turbulent fluctuations has yet to be well established.

In plasmas, the turbulent cascade initiates at large spatial scales and energy is nonlinearly transferred toward kinetic scales, where it is finally dissipated. At proton characteristic scales, on the order of the Larmor radius ($\rho_i$) and the ion inertial length ($d_i$), the fluctuations develop a strong compressive component and wave activity. These fluctuations have been identified as kinetic Alfv\'en waves (KAWs) around $k_\perp \rho_i \gtrsim 1$ \citep{leamon1998observational,sahraoui2009evidence,salem2012identification,kiyani2013enhanced,chen2013nature}. Their role in the dissipation range has been extensively investigated through theoretical analyses and numerical simulations \citep{hollweg1999kinetic,howes2008model,schekochihin2009astrophysical,schekochihin2022mhd,gary2004kinetic,vasconez2015kinetic,franci2015high,cerri2016subproton,grovselj2018fully,grovselj2019kinetic,zhou2023electron}. Moreover, there is also substantial evidence of the presence of circularly polarized waves at frequencies near the proton gyrofrequency, commonly identified as ion cyclotron waves (ICWs) and fast magnetosonic waves (FMWs)  \citep{jian2009ion,jian2014electromagnetic,bowen2020ion,zhao2021mhd,liu2023radial}.

Ion-scale waves could undergo resonant wave–particle interactions, including Landau damping \citep{leamon1999dissipation,sahraoui2010three,kobayashi2017three}, transit-time damping  \citep{barnes1966collisionless,stix1992waves}, and cyclotron resonance  \citep{hollweg2002generation,isenberg2011kinetic,isenberg2019perpendicular,johnston2026quasi}. In addition, non-resonant processes—such as stochastic heating driven by large-amplitude fluctuations at the proton gyroscale \citep{chandran2010perpendicular,hoppock2018stochastic}, together with the intermittent nature of turbulent fluctuations and the associated coherent structures, are also expected to make a substantial contribution to particle energization \citep{dmitruk2004test,greco2009statistical,osman2012kinetic,mallet2019interplay,mallet2026perpendicular,2025PhRvL.135y5201B}. These mechanisms may operate simultaneously or be activated under particular plasma conditions, giving rise to different heating channels and signatures in velocity space \citep{verniero2022strong,gonzalez2024local}. 

Non-Maxwellian velocity distribution functions (VDFs) are routinely observed in the solar wind, and these VDFs often exhibit drifting particle populations and temperature anisotropies that provide free energy for the excitation of kinetic instabilities. The resulting instabilities act to regulate the degree of non-thermal structure in the VDFs and constitute an important source of ion-scale waves \citep{gary1993theory,Daughton_Gary98}. A growing body of observational evidence indicates that ion-scale waves in the inner heliosphere are predominantly driven by proton temperature anisotropies and proton beam populations \citep{woodham2019parallel,Verniero_2020,klein2021inferred,liu2025effects,niranjana2026free}.

A key parameter in solar wind turbulence is the normalized cross-helicity, which measures the correlation between velocity and magnetic field fluctuations. The cross-helicity is an invariant in the incompressible ideal MHD framework, and measures the imbalance between AWs propagating parallel/antiparallel to the mean field. As such, it controls the strength of nonlinear interaction and the available energy in the cascade \citep{iroshnikov1963turbulence,kraichnan1965inertial}. Extended models showed that strong imbalanced turbulence (with high cross-helicity) is subject to the ``helicity barrier'' \citep{meyrand2021violation}---an effect that results as a consequence of the conservation of energy and generalized-helicity in low-$\beta$ gyrokinetics \citep{schekochihin2019constraints,chandran2026second}---in which the energy gets stuck at $k_\perp\rho_i \lesssim 1$, producing a steep transition in the spectrum at sub-proton scales, and the building up energy in the inertial range. A fraction of the energy is able to transport into small parallel scales ($k_\parallel d_i \sim 1$) in the form of ICWs, which is likely generated by a secondary instability triggered by oblique cyclotron resonant heating \citep{chandran2010,squire2022high}. Evidence for the helicity barrier has been found in the transition range scaling \citep{mcintyre2025} and additionally ion scale waves have been found to be statistically correlated with intervals of imbalanced turbulence.\citep{bowen2024mediation,zhao2026ion}. These quasi-parallel ICWs heat protons \citep{squire2023electron} as well as minor ion species \citep{zhang2025extreme}. On the other hand, the barrier is absent in balanced turbulence, allowing the cascade to reach kinetic scales and enabling development of intermittency that heat protons via stochastic heating and resonant interactions \citep{arzamasskiy2019hybrid,cerri2021stochastic}.

The imbalanced turbulence studies mentioned in the previous paragraph all relied on driven simulations which, in combination with the helicity barrier, cause energy to build up in the inertial range. This accumulation can promote an inverse cascade that may subsequently influence particle dynamics\citep{gonzalez2017test}. Although continuous driving is physically relevant to the solar corona and solar wind because the Sun continuously injects outward-propagating waves, the cascade is believed to be initiated nonlinearly via wave reflection \citep{velli1989turbulent,chandran2009alfven,chandran2019reflection,meyrand2025reflection}. To investigate this further, we focus on a complementary idealization to the forced turbulence studies, freely decaying balanced and imbalanced turbulence. The goal of this work is to investigate kinetic-scale fluctuations and energy dissipation across different turbulent regimes using 3D hybrid-kinetic simulations. We focused on quantifying proton heating, the partitioning between parallel and perpendicular temperature, and their connection with wave activity driven by kinetic instabilities. This paper is organized as follows: Section~\ref{datasec} outlines the hybrid-kinetic model and the numerical configuration. Section~\ref{res} presents the time evolution of the simulations and heating diagnostics; it also shows the spectral properties at large and sub-proton scales for each case, and then examines ion-scale wave generation and phase-space dynamics in both balanced and imbalanced turbulence. Finally, Section~\ref{coclusions} summarizes the main findings and discusses their implications.

\section{Model}
\label{datasec}

We considered a quasi-neutral hybrid-kinetic plasma model in which electrons are described as a massless and isothermal fluid, while protons are treated kinetically using the particle-in-cell method. The simulations were performed with CAMELIA code \footnote{https://space.asu.cas.cz/~helinger/camelia.html}, that is based on current advanced method and cyclic leapfrog \citep{matthews1994,Franci_2018}. The code uses the following normalization scheme: time is normalized to the inverse proton gyrofrequency, $\Omega_{ci}^{-1} = (eB_0/m_i c)^{-1}$. Lengths are expressed in terms of the ion inertial length $d_i = c/\omega_{pi}$, where the plasma frequency is $\omega_{pi} = (e^2 n_0 c^2 \mu_0/m_i)^{1/2}$. Velocities are scaled to the Alfvén speed $v_A = B_0/\sqrt{\mu_0 n_0 m_i}$, with $\mu_0$ denoting the vacuum magnetic permeability and $n_0$ the initial plasma density..

The simulation setup consists of a uniform background magnetic field $\mathbf{B}_0$ along the \textit{x}-direction, and we introduced perturbations in the transverse components of the magnetic field ($\delta \mathbf{b}/B_0$) and proton bulk velocity ($\delta \mathbf{u}/v_A$). Fluctuations are initialized with an isotropic broadband energy spectrum $E_{\mathbf{u},\mathbf{b}}(\mathbf{k}) \propto |\mathbf{k}|^{-1}$ in the range $ 0.0245 \leq|\mathbf{k}|\leq 0.147$, and it is constructed by the superposition of wave modes of equal amplitude and random phase. We examined two cases, differentiated by their initial normalized cross-helicity $
\sigma_c = 2 {\langle {\bf u}\cdot{\bf b^*}\rangle}/{\langle |{\bf u}|^2\rangle + \langle |{\bf b^*}|^2\rangle}$; here $\mathbf{b^*} =  \mathbf{b}/\sqrt{n_0}$ is expressed in units of $v_A$. The first corresponds to a relatively balanced turbulence with initial $\langle \sigma_c \rangle = -0.25$, while the second is highly imbalanced, with $\langle \sigma_c \rangle = -0.98$. The fluctuations are in equipartition between kinetic and magnetic energy, that is quantified through the normalized residual energy $\sigma_r = ({|\mathbf{u}|^2 - |\mathbf{b^*}|^2})/({|\mathbf{u}|^2 + |\mathbf{b^*}|^2})$, with initial $\langle \sigma_r \rangle = 0$ in both cases. 

We perform 3D simulations in a periodic box of size $256 d_i \times128^2 d_i $, with a uniform grid spacing of $\Delta  = 0.5\, d_i$ and a time step of $\Delta t = 0.05\, \omega_{ci}^{-1}$. We used 512 particles-per-cell, for a total of $1.71\times 10^{10}$ particles. The initial fluctuation amplitude is  $\delta b/B_0 = \delta u/v_A = 0.489$. The proton velocities were initialized using a shifted-Maxwellian distribution, and the plasma beta for both protons and electrons is set to $\beta_{p,e} = {2\mu_0n_0T_{p,e}}/{B_0^2}=0.5$. To control numerical dissipation at the grid scale, we include an explicit resistivity of $\eta  = 1.5 \times 10^{-3} \mu_0 v_A^2 \omega_{ci}^{-1}$.

\section{Results}
\label{res}

\subsection{Overview:}

\begin{figure}[ht!]
\includegraphics[width=0.49\textwidth]{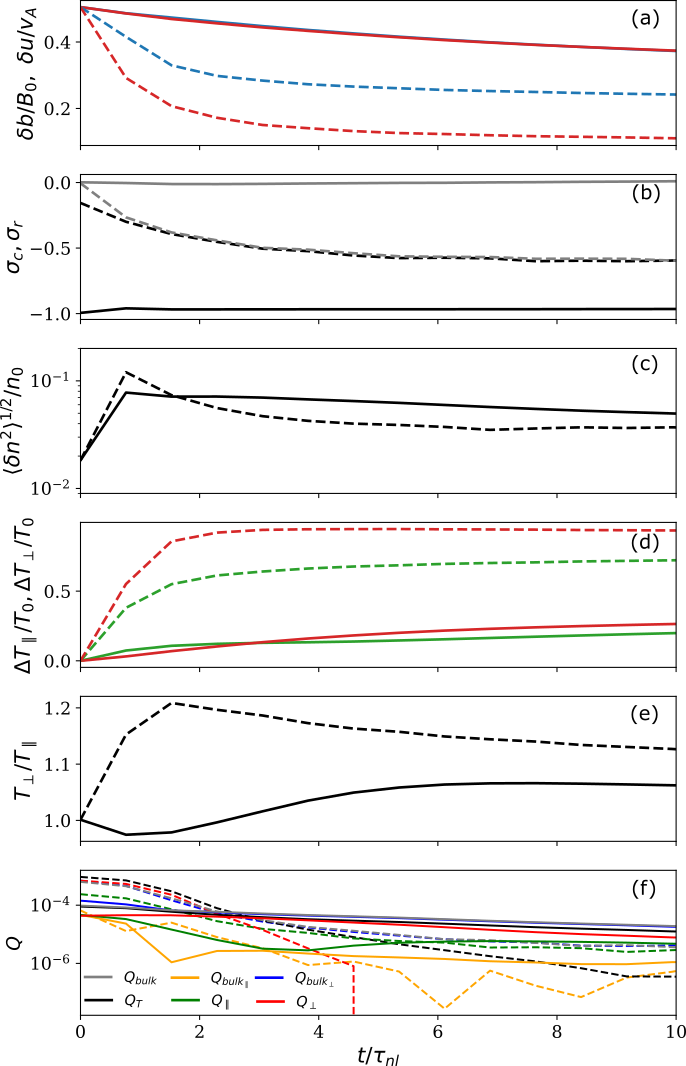}
\caption{Time evolution for the imbalanced (solid curves) and balanced simulations (dashed curves). (a) Normalized amplitudes of magnetic (blue) and velocity (red) fluctuations; (b) mean cross-helicity (black) and residual energy (gray); (c) the rms of density fluctuations; (d) the changes in parallel (green) and perpendicular proton temperature ((red) to the mean field; (e) the average temperature anisotropy $T_\perp/T_\parallel$; (f) heating and dissipation rates.} 
\label{fig:zero}
\end{figure}

\begin{figure*}[t!]
\includegraphics[width=0.99\textwidth]{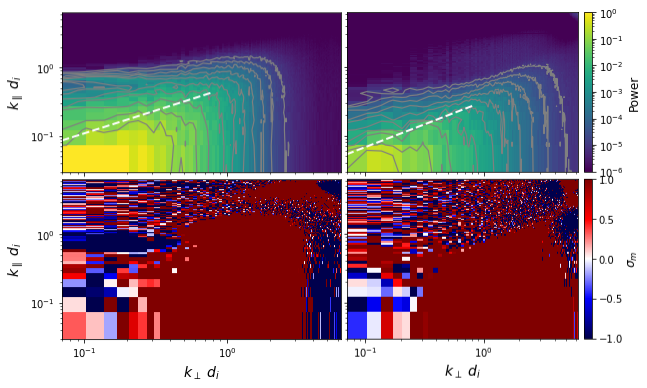}
  \caption{The 2D PSD for $|\delta B_\perp(k_\perp,k_\parallel)|^2$ at $t = 4.6 \tau_{nl}$ (top panels) and the magnetic helicity $\sigma_m(k_\perp,k_\parallel)$ (bottom panels) for the imbalanced (left) and balanced simulation (right). }
\label{fig:two}
\end{figure*}

Fig.~\ref{fig:zero} presents the global temporal evolution of the simulations. The results are shown in terms of the nonlinear time $\tau_{nl} = L_\perp/\delta b$, with $L_\perp$ the box size perpendicular to the guide magnetic field. In the imbalanced case (solid lines), fluctuations propagate along the background magnetic field. At early times ($t < 1\ \tau_{nl} \equiv 261.75 \ \omega_{ci}^{-1}$), the system rapidly develops nonlinear compressible fluctuations driven by variations in $n$ and $|\mathbf{B}|^2$ (not shown), which, through their coupling with Alfvénic fluctuations, result in the steepening of the fluctuations \cite[see, e.g.,][]{machida_simulation_1987,buti2000hybrid,Gonzalez_2021,bianco2026evolution}. These compressible fluctuations, in turn, generate wave reflection, which produces a modest reduction in $\sigma_c$ and removal of energy from the fluctuations, as evidenced by the decrease in $\delta b/B_0$ and $\delta u/v_A$ (top panel). On the hand, in the balanced case (dashed lines), large scale eddies develop and the fluctuation's energy is efficiently transported and removed from the bulk resulting in stronger heating (see fourth panel). For that simulation, kinetic energy is dissipated more effectively than magnetic energy, with average values of $\langle \sigma_c \rangle \sim -0.4$ and $\langle \sigma_r \rangle \sim -0.6$ once the system reaches the fully developed turbulent state at approximately $2\tau_{nl}$.

We evaluated the normalized temporal variation of the proton temperature relative to its initial value, defined as $\Delta T_{\parallel,\perp} = {(T_{\parallel,\perp} - T_{0_{\parallel,\perp}}})/{T_{0_{\parallel,\perp}}}\,, $ with \(T_{0_{\parallel,\perp}}\) the initial temperature. Here we define parallel and perpendicular temperature as $T_\parallel = \textbf{p} \colon \hat{\textbf{b}}\hat{\textbf{b}}/n$ and $T_\perp = \textbf{p} \colon (\mathbb{\textbf{I}} - \hat{\textbf{b}}\hat{\textbf{b}})/2 n$, with the pressure tensor $\textbf{p} = \int(u-v_p)_i(u-v_p)_j f({\bf r,v},t) d{\bf v}$ obtained from the particle velocity distribution and $\hat{\textbf{b}} = \textbf{B}/\Vert \textbf{B} \Vert$ the direction of the total magnetic field.

Variations in proton temperature differ between the balanced and imbalanced simulations. Although these variations are primarily observed in $T_\perp$ in both cases, each run also exhibits an increase of $T_\parallel$, albeit at different stages in the evolution (fourth panel). In the balanced case, the temperature anisotropy initially displays $T_\perp>T_\parallel$, which gradually decreases toward $T_\perp/T_\parallel\sim 1.1$ as the main heating process reaches saturation. This evolution is accompanied by the mild increase of $T_\parallel$ at later times. In the imbalanced simulation, $T_\parallel$ exhibits a pronounced initial increase, resulting in $T_\perp/T_\parallel<1$ during the first turnover time. This increase is due to acceleration of protons along the magnetic field and the generation of a proton beam (discussed below). That stage is subsequently followed by a persistent, nearly steady, secular growth of $T_\perp$ (accompanied also by a comparatively weaker increase in $T_\parallel$), which continues up to the end of the simulation.

The bottom panel in Fig.~\ref{fig:zero} shows the heating rate $Q_T = \langle { \frac{\partial}{\partial t}[\frac{1}{2}  n (T_\parallel + 2 T_\perp)}] \rangle$, which quantifies the overall rate of plasma temperature increase. In the same way, we calculate the parallel and perpendicular heating rates, $Q_{\parallel} =  \,\langle {\frac{\partial}{\partial t} (\frac{1}{2} n T_{\parallel})}\rangle$ and $Q_{\perp} = \langle {\frac{\partial}{\partial t}( nT_{\perp})}\rangle$, such that $Q_T = Q_\parallel+ Q_\perp$. Furthermore, we evaluate the rate of loss of bulk energy, $Q_{\rm bulk} = - \,\langle{\frac{\partial}{\partial t} \left( \frac{1}{2} n (\delta b^2 + \delta u^2)\right)}\rangle$,  which measures the rate at which energy is removed from magnetic and velocity fluctuations. In addition, we compute $Q_{\perp_{\rm bulk}} = - \,\langle \frac{\partial}{\partial t}  \left( \frac{1}{2} n(\delta b_\perp^2 + \delta u_\perp^2) \right)\rangle$ and $Q_{\parallel_{\rm bulk}} = -\,\langle {\frac{\partial}{\partial t} \left( \frac{1}{2} n(\delta b_\parallel^2 + \delta u_\parallel^2)\right)} \rangle$, representing the transverse and parallel bulk energy contributions, respectively. Here $\delta b$ is expressed in Alfven units, and we note that for the bulk energy rate, we refer to parallel/perpendicular to $\mathbf{B}_0$

The proton heating rate is approximately an order of magnitude larger in the balanced than in the imbalanced simulation, owing to the  counter-propagating fluctuations that trigger a more rapid and intense onset of the turbulent cascade at early times. During this initial transition toward a fully developed turbulent state, the proton heating is predominantly perpendicular, and the energy gained by the protons is, to a good approximation, comparable to the energy extracted from the bulk plasma (as indicated by the black and the gray curves in the bottom panel). Once turbulence becomes fully developed, the perpendicular heating rate reaches a saturation level, followed by a rapid drop in the perpendicular heating rate. For $t\gtrsim3\tau_{nl}$, the energy extracted from the bulk is instead converted primarily into parallel heating, with a reduced heating rate compared to the initial nonlinear stage. In the imbalanced case, not only the heating rate is lower, but the overall heating process differs significantly from the balanced simulation. At early times, there is parallel heating due to the phase-steepening of the fluctuations that a portion of the bulk energy is converted into the acceleration of protons that form the beam. Following this initial stage ($t\gtrsim\tau_{nl}$), the net heating in the imbalanced simulation becomes predominantly perpendicular (with $Q_\perp/Q_\parallel \sim 10$), and the heating rate decreases only slowly over time.

\subsection{Turbulent properties and ion-scale wave generation:}

\begin{figure}[htbp]
\includegraphics[width=0.45\textwidth]{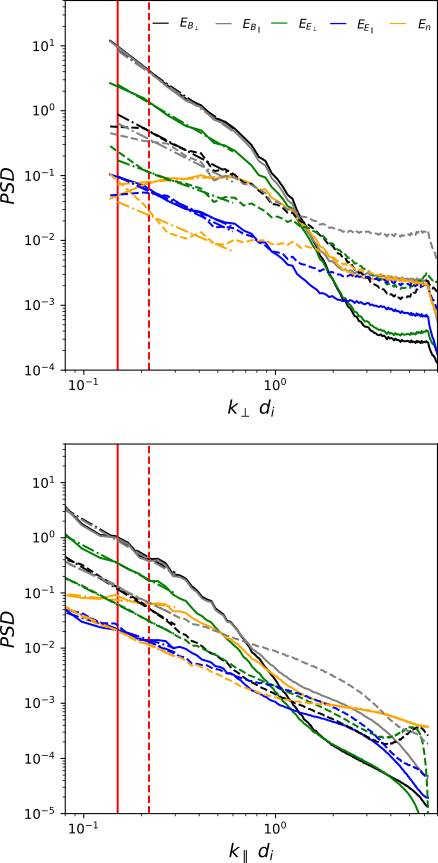}
  \caption{The reduced spectra of  $E_{B_\perp}$ (black), $E_{B_\parallel}$ (gray), $E_{E_\perp}$ (green), $E_{E_\parallel}$ (blue), and $E_{\rho}$ (orange) for the imbalanced (solid lines) and balanced (dashed lines) turbulence simulations. The top panel displays the reduced spectra as functions of the perpendicular wavenumber $k_\perp$, while the bottom panel shows the spectra along the parallel direction. The spectrum is plotted at $t = 4.6\,\tau_{nl}$. The solid and dashed red lines represent the wavevector associated to the energy containing scale $L_{0}$ for the balanced and imbalanced simulations, respectively.}
\label{fig:four}
\end{figure} 

Fig \ref{fig:two} top panel presents the 2D power spectral density (PSD) of the transverse magnetic field fluctuations $|\delta B_\perp(k_\perp,k_\parallel)|^2$ at $t=4.6 \ \tau_{nl}$. The bottom panels display the normalized magnetic helicity, $\sigma_m = k H_k/E_{B_\perp} $, with $H_k = i(B_y^*B_z - B_z^*B_y)/k_x$ and $E_{B_\perp} = (B_y^2 + B_z^2)$, quantifying the handedness of the fluctuations at different scales \citep{howes2010interpretation}. We found that in both simulations, energy is concentrated at large scales and the cascade proceeds toward smaller scales following close critical balance ($k_\parallel \propto k_\perp^{2/3}$)\citep{goldreich1995toward}. However, the spectral properties at large and kinetic scales differ significantly in the simulations. At large scales, fluctuations contain contributions from multiple modes, with $|\langle \sigma_m|\rangle \sim 0$. Moving toward smaller scales, the fluctuations exhibit a clear right-handed polarization ($\sigma_m \sim 1$), that might result from the transition toward KAWs at perpendicular sub-proton scales. This is also evident in the imbalanced simulation, which also shows a preferred right-handed polarization at perpendicular kinetic scales. However, the imbalanced turbulence simulation develop a distinct region around $k_\parallel d_i \sim 1$, with substantial power along the parallel direction that displays left-handed polarization ($\sigma_m=-1$), consistent with the presence of ion-cyclotron waves. The spectral features developed in the imbalanced case is consistent with previous reported imbalanced simulations \citep{squire2022high,markovskii2023magnetic,zhang2025extreme} as well as observations of the solar wind \citep{podesta2011magnetic,he2011possible,huang2020kinetic}.

Fig.~\ref{fig:four} presents the perpendicular and parallel reduced PSD for several fields in both the balanced and imbalanced simulations at $t = 4.6\tau_{nl}$. In particular, for the perpendicular and parallel components of the magnetic and electric fields, together with the spectra of plasma density. Although the total energy differs between the two simulations at this specific time, their spectral properties remain qualitatively similar throughout the late stages of the evolution. In the balanced case, the perpendicular magnetic energy spectrum exhibits a power-law scaling $E_{B_\perp}(k_\perp) \propto k_\perp^{-5/3}$ at large scales and steepens to $E_{B_\perp}(k_\perp) \propto  k_\perp^{-8/3}$ at sub-proton scales. Along the parallel direction, we obtain $E_{B_\perp}(k_\parallel) \propto k_\parallel^{-2}$ across both large and kinetic scales. The perpendicular electric field spectrum displays a slope $E_{E_\perp}(k_\perp) \propto k_\perp^{-1.5}$ in the inertial range, which flattens at sub-proton scales to approximately $E_{E_\perp}(k_\perp) \propto  k_\perp^{-1.0}$. These scalings are in close agreement with the spectral transitions reported in previous balanced turbulence simulations \citep{Franci_2018,cerri2019kinetic,grovselj2018fully,gonzalez2019turbulent}. 

In contrast, in the imbalanced simulation the magnetic spectrum scales as $E_{B_\perp}(k_\perp)\propto k_\perp^{-2}$ and $E_{B_\parallel}(k_\perp) \propto  k_\perp^{-2}$ over the inertial range, following weak turbulence prediction \citep{galtier2000weak}. At sub-proton scales, however, the perpendicular magnetic spectrum exhibits a step transition, with $E_{B_\perp}(k_\perp) \propto  k_\perp^{-4.9}$, in agreement with previous numerical investigations of imbalanced turbulence \citep{meyrand2021violation,squire2022high,squire2023electron}. Remarkably, the compressive contribution from density and parallel magnetic fluctuations dominate sub-proton scales in both $k_\perp$ and $k_\parallel$. The imbalanced simulation shows stronger density power near the transition region compared to the balanced run,  and this might be the result of compressible fluctuations avoiding the helicity barrier. In contrast, the balanced case exhibits larger parallel magnetic and electric field at proton scales, that is the results of the intermittent current sheets and reconnecting fields inside those small-scale structures \citep{dmitruk2006structure,sahraoui2010three,bowen2020constraining}.

\begin{figure}[ht!]
\includegraphics[width=0.49\textwidth]{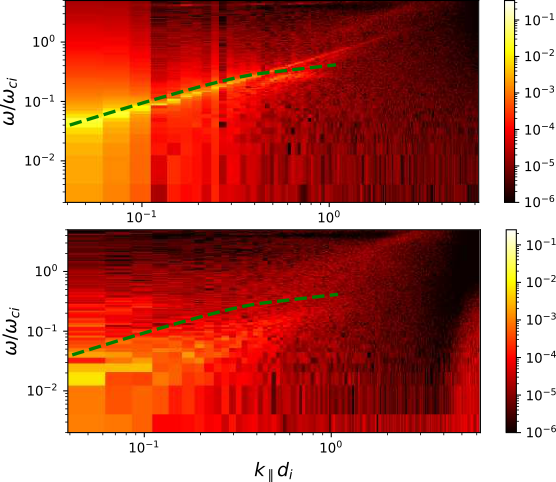}
  \caption{The $k_\parallel-\omega$ spectrum for imbalanced (top) and balanced turbulence simulations (bottom). The green dashed line correspond to linear dispersion relation for ICWs in the hybrid-kinetic model.}
\label{fig:three}
\end{figure}

In addition, we investigated whether an inverse cascade develops in the imbalanced run by examining the time evolution of the energy-containing scale (not shown), that is defined as $L_{0} = \int{E_{B_\perp}(k) dk}/\int{ k^{-1}E_{B_\perp}(k) dk}$. We plotted the corresponding wave number at at $t = 4.6\tau_{nl}$, represented by the red vertical lines in Fig~\ref{fig:three}. While the simulation displays a step transition at sub-proton scales, which may be compatible with the helicity barrier, we find no indication that $L_0$ increases in our simulation.

\begin{figure*}[ht!]
\includegraphics[width=0.95\textwidth]{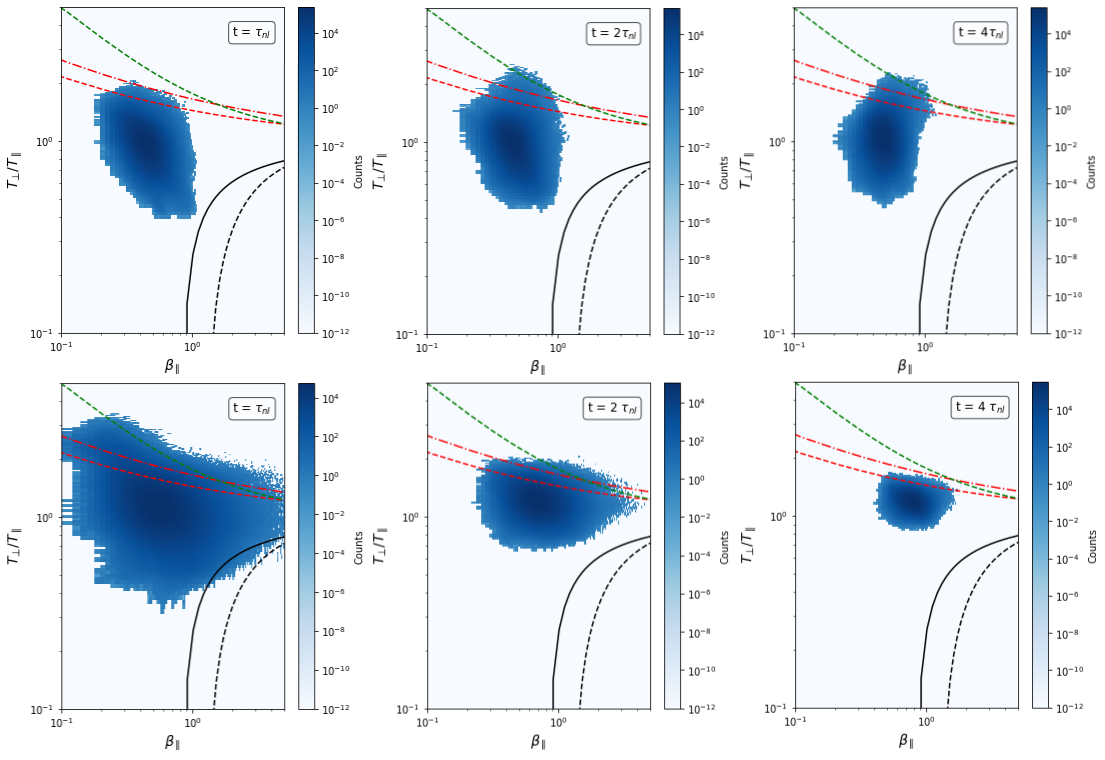}
  \caption{ Representation of the simulation data in the ($\beta_\parallel, T_\perp/T_\parallel$) plane at several times for the imbalanced (top) and balanced (bottom) turbulence simulations. The curves represent the instability thresholds predicted by linear theory for various kinetic instabilities. Black lines indicate the parallel (solid) and oblique (dashed) firehose instabilities for maximum growth rate $\gamma/\omega_{ci} = 10^{-3}$. The green and red lines correspond to the mirror and ICW instabilities, respectively. For the ICW instability, we show predictions for growth rates $\gamma/\omega_{ci} = 10^{-3}$ (red dashed) and $\gamma/\omega_{ci} = 10^{-2}$ (red dash-dotted).}
\label{fig:six}
\end{figure*}

\begin{figure*}[ht!]
\includegraphics[width=0.95\textwidth]{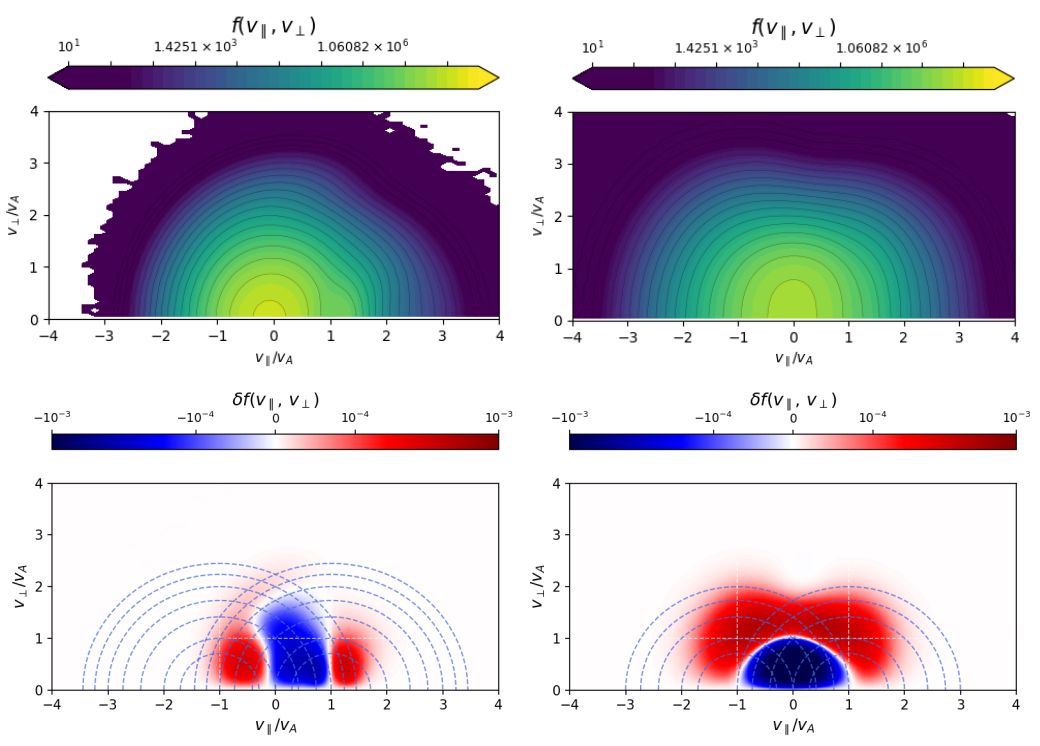}
  \caption{The reduced velocity distribution function $f(v_\parallel,v_\perp)$ (top) and the variations of the VDF $\delta f(v_\parallel,v_\perp)$ (bottom) at $t = 5 \tau_{nl}$ for the imbalanced (left) and balanced turbulence simulations (right).}
\label{fig:seven}
\end{figure*}

Fig \ref{fig:three} shows the $k_\parallel-\omega$ spectrum for the imbalanced and balanced simulations. This was obtained by taking the magnetic field for over $500 \ \omega_{ci}^{-1}$ at a cadence $0.2 \omega_{ci}^{-1}$ in the time range $t=\{7.6-9.5\} \tau_{nl}$. Then, time-space data was Fourier transformed to obtain the spectrogram $|\delta B_\perp(k_{\parallel},\omega)|^2$. We also plotted the dispersion relation of ICWs in the Vlasov-Maxwell hybrid model, that was obtained using the HYDROS code \citep{told2016linear}. In the imbalanced turbulence case, the parallel-propagating ICWs closely follow the linear dispersion relation according to the hybrid kinetic model. By contrast, the balanced simulation displays low-amplitude fluctuations below the ICW branch, with only significant amplitude in the low frequency region corresponding to the Alfven branch. Although some features are observed near $k_\parallel d_i \sim 1$, fluctuation's amplitude remain sufficiently small that they do not seem to contribute significantly to the energetics.

The ICWs activity that develops in the simulation is linked to unstable proton VDFs, which can be analyzed in the parameter space defined by the parallel proton beta and the temperature anisotropy plane ($\beta_\parallel, T_\perp/T_\parallel$). Figure~\ref{fig:six} shows the time evolution of the simulations in that plane and the curves indicate predictions for the onset of different plasma instabilities with a growth rate $\gamma/\omega_{ci} = 10^{-3}$, delimiting the thresholds for the parallel and oblique firehose instabilities (solid and dashed black lines), as well as for the mirror (green) and ion-cyclotron instabilities (red) \citep{hellinger2006solar}. We should mentioned that these thresholds are defined based on the assumption of bi-Maxwellian distribution, which may not be correct here. In the imbalanced simulation, a small fraction of data points exceed the instability threshold for ICW instability. This is consistent with the observed left-handed polarized fluctuations at $k_\parallel d_i\sim1$ that follow the ICW dispersion relation (Fig~\ref{fig:two} and Fig~\ref{fig:three}). These waves are generated in regions with sufficient temperature anisotropy, providing enough free energy to excite ICWs. The unstable portion persists over long times in the simulation, which is also consistent with the sustained perpendicular heating observed in the imbalanced case. For the balanced simulation, during the early stage when turbulence is still developing (at $t=\tau_{nl}$), the plane is densely populated over a wide range of $\beta_\parallel$ and $T_\perp/T_\parallel$, but remain constrained near the marginal stability threshold. However, once the system reaches a fully developed turbulent state, the temperature anisotropy relaxes and plasma shifts into a stable region bounded by instabilities thresholds.

\subsection{Phase space dynamics}

Fig~\ref{fig:seven} shows the proton VDF and the departure from Maxwellian distribution (measured through the changes from the initial distribution, $\delta f(v_\parallel,v_\perp) = f - f_0$. In the imbalanced case, a proton beam forms at the beginning of the simulation, and remains present throughout its long-term evolution. This beam is generated by the acceleration of protons during the steepening of the Alfvénic fluctuations. The velocity-space structure in the imbalanced run highlights the importance of resonant processes, with strong activity near $v \sim v_A$ signature of particle acceleration into the beam. Another feature is the scattering of protons along the diffusion contours from ICWs. The blue dashed curves show the trajectory of a particle whose energy is conserved in the wave frame $(v_\parallel - \omega/k_\parallel)^2 + v_\perp^2 = I$, assuming the phase-velocity of the wave $\omega/k_\parallel\sim \pm v_A$\citep{kennel1966velocity}. It is noticed the signatures of particles with negative velocities interacting with forward ICWs. Also, protons on the core with positive velocities seems to be scattered by backward propagating waves, in which the blue shade region is bended toward the left. Therefore, we can argue from the velocity-space signatures that the  extended perpendicular heating in the imbalanced simulation is the result of  resonant interactions with ICWs.

On the other hand, in the balanced simulation, VDF also develops non-Maxwellian features, but they do not follow dispersion contours as in the imbalanced case. Specifically, the core of the distribution evidences the combined influence of Landau damping—seen as particle bumps around $\pm v_{th}$—and the emergence of two lobes populated with a substantial number of protons with enhanced $v_\perp$, which are displaced from and energized relative to the core. This velocity-space structure remains stable once the simulation undergoes strong perpendicular heating during the initial transition to fully developed turbulence (up to $2 \ t/t_{nl}$). Such energization might result from the combination of stochastic and resonant heating mechanisms, as reported in previous 3D hybrid simulations of balanced turbulence \citep{cerri2021stochastic}.

\section{Summary}
\label{coclusions}

In this work, we use three-dimensional hybrid-kinetic simulations to examine kinetic-scale wave activity and proton heating in Alfvénic balanced and imbalanced turbulence. We study a freely decaying turbulence configuration with identical initial perturbations and fluctuation amplitudes, but with different correlations between magnetic and velocity fluctuations (quantified by $\sigma_c$), which correspond to intrinsically different turbulent regimes. The balanced case exhibits a more efficient nonlinear cascade and a substantially higher dissipation/heating rate, accompanied by a faster removal of bulk fluctuation's energy during the evolution toward a fully developed turbulent state. In both regimes, sub-proton scales are dominated by highly oblique, right-hand-polarized compressive fluctuations with properties of KAWs \citep{franci2015high,cerri2016subproton,grovselj2018fully,grovselj2019kinetic}. In addition, the imbalanced simulation displays enhanced power at finite parallel wavenumbers, $k_\parallel d_i \sim 1$, consistent with ICWs, in agreement with previous imbalanced turbulence simulations \citep{squire2022high,squire2023electron,markovskii2023magnetic,zhang2025extreme}.

The distinct kinetic activity in the simulations is associated with different proton energization processes. In the balanced case, the system exhibits pronounced early-time perpendicular heating, which is subsequently followed by enhanced parallel heating at later times. In contrast, the imbalanced simulation undergoes initial parallel heating, resulting from the acceleration of protons and beam formation, and followed by stronger perpendicular heating connected to ion-cyclotron heating. In particular, the imbalanced run consistently exhibits unstable temperature anisotropy near ion-cyclotron instability and signatures of proton scattering by ion-scale waves. The ICWs generation is associated to local temperature anisotropy at phase-steepened fronts that is driven by the core proton population, similarly to processes studied in hybrid simulations in a simplified geometry  \citep{Gonzalez_2021,Qian_2026}. The velocity-space diagnostics support these interpretations: in the early stage, the imbalanced simulation undergoes phase steepening that accelerates protons to velocities comparable to the Alfvén speed and generates complex kinetic dynamics indicative of beam formation (at positive velocities). Additionally, phase-space show signature of scattering by ICWs that results in extended heating once ion-scales waves are generated. On the other hand, the balanced simulation also exhibits distortions of the core velocity distribution, but non-Maxwellian features are different with  enhanced $v_\perp$ populations that are consistent with the interplay of resonant and non–resonant heating mechanisms \citep{arzamasskiy2019hybrid,cerri2021stochastic}.

These results are directly relevant to the solar wind, particularly in the inner heliosphere, where in-situ measurements have revealed both strongly imbalanced turbulence and a prominent population of ICWs \citep{bowen2024mediation,niranjana2026free,zhao2026ion}. Our simulations provide numerical support for the pathway in which the nonlinear evolution of imbalanced Alfv\'enic fluctuations leads to the generation of ICWs and enables efficient proton heating through cyclotron resonant under conditions of the near-Sun solar wind and solar corona. At the same time, it has been argued that cyclotron resonance alone may be insufficient to supply the required heating in the sub-Alfv\'enic wind, and that non-resonant mechanisms, in particular stochastic heating, may be more viable mechanism \citep{bowen2025stochastic}. In our imbalanced simulation, we do not observe clear evidence for strong stochastic heating; instead, the dominant energization is tied to the steepening of Alfv\'enic fluctuations and subsequent resonant interactions with ICWs. A quantitative comparison of resonant and non-resonant contributions will require estimates of fully kinetic heating rate in both balanced and imbalanced regimes \citep{ klein2016measuring,ervin2026direct}. More broadly, these results highlight that nonlinear dynamics and turbulent heating depend sensitively on the degree of Alfv\'enicity of the plasma. Future investigations should extend the present analysis across a wider range of plasma parameters (e.g., plasma $\beta$ and fluctuation amplitudes) and incorporate additional physics relevant to the solar wind, including forcing, large-scale expansion and electron kinetic effects, to better constrain energy partition and dissipation and their role in plasma heating and acceleration in heliospheric and other astrophysical environments.

\begin{acknowledgments}
This research was supported by NSF SHINE grant \#80NSS\-C18K1211. We acknowledge the Texas Advanced Computing Center (TACC) at The University of Texas at Austin for providing HPC resources for this investigation. The simulations have been run on the Frontera and Lonestar6 supercomputer http://www.tacc.utexas.edu.
\end{acknowledgments}

\bibliography{sample7}

@article{bruno2013solar,
  title={The solar wind as a turbulence laboratory},
  author={Bruno, Roberto and Carbone, Vincenzo},
  journal={Living Reviews in Solar Physics},
  volume={10},
  number={1},
  pages={1--208},
  year={2013},
  publisher={Springer}
}

@article{matthews1994,
  title={Current advance method and cyclic leapfrog for 2D multispecies hybrid plasma simulations},
  author={Matthews, Alan P},
  journal={Journal of Computational Physics},
  volume={112},
  number={1},
  pages={102--116},
  year={1994},
  publisher={Elsevier}
}

@article{Franci_2018,
	year = 2018,
	month = {may},
	publisher = {{IOP} Publishing},
	volume = {1031},
	pages = {012002},
	author = {L. Franci and P. Hellinger and M. Guarrasi and C. H. K. Chen and E. Papini and A. Verdini and L. Matteini and S. Landi},
	title = {Three-dimensional simulations of solar wind turbulence with the hybrid code {CAMELIA}},
	journal = {Journal of Physics: Conference Series}
}

@article{franci2015high,
  title={High-resolution hybrid simulations of kinetic plasma turbulence at proton scales},
  author={Franci, Luca and Landi, Simone and Matteini, Lorenzo and Verdini, Andrea and Hellinger, Petr},
  journal={The Astrophysical Journal},
  volume={812},
  number={1},
  pages={21},
  year={2015},
  publisher={IOP Publishing}
}

@article{Gonzalez_2021,
	year = 2021,
	month = {jun},
	publisher = {American Astronomical Society},
	volume = {914},
	number = {2},
	pages = {L36},
	author = {C. A. Gonz{\'{a}}lez and A. Tenerani and L. Matteini and P. Hellinger and M. Velli},
	title = {Proton Energization by Phase Steepening of Parallel-propagating Alfv{\'{e}}nic Fluctuations},
	journal = {The Astrophysical Journal Letters}
}

@article{Verniero_2020,
year = 2020,
month = {apr},
publisher = {American Astronomical Society},
volume = {248},
number = {1},
pages = {5},
author = {J. L. Verniero and D. E. Larson and R. Livi and A. Rahmati and M. D. McManus and P. Sharma Pyakurel and K. G. Klein and T. A. Bowen and J. W. Bonnell and B. L. Alterman and P. L. Whittlesey and David M. Malaspina and S. D. Bale and J. C. Kasper and A. W. Case and K. Goetz and P. R. Harvey and K. E. Korreck and R. J. MacDowall and M. Pulupa and M. L. Stevens and T. Dudok de Wit},
title = {Parker Solar Probe Observations of Proton Beams Simultaneous with Ion-scale Waves},
journal = {The Astrophysical Journal Supplement Series}
}

@article{Daughton_Gary98,
author = {Daughton, William and Gary, S. Peter},
title = {Electromagnetic proton/proton instabilities in the solar wind},
journal = {Journal of Geophysical Research: Space Physics},
volume = {103},
number = {A9},
pages = {20613-20620},
eprint = {https://agupubs.onlinelibrary.wiley.com/doi/pdf/10.1029/98JA01385},
year = {1998}
}

@article{cerri2021stochastic,
  title={On stochastic heating and its phase-space signatures in low-beta kinetic turbulence},
  author={Cerri, Silvio Sergio and Arzamasskiy, Lev and Kunz, Matthew W},
  journal={The Astrophysical Journal},
  volume={916},
  number={2},
  pages={120},
  year={2021},
  publisher={IOP Publishing}
}

@article{verniero2022strong,
  title={Strong perpendicular velocity-space diffusion in proton beams observed by parker solar probe},
  author={Verniero, JL and Chandran, BDG and Larson, DE and Paulson, K and Alterman, BL and Badman, S and Bale, SD and Bonnell, JW and Bowen, TA and de Wit, T Dudok and others},
  journal={The Astrophysical Journal},
  volume={924},
  number={2},
  pages={112},
  year={2022},
  publisher={IOP Publishing}
}

@article{tu1995mhd,
  title={MHD structures, waves and turbulence in the solar wind: Observations and theories},
  author={Tu, C-Y and Marsch, Eckart},
  journal={Space Science Reviews},
  volume={73},
  number={1},
  pages={1--210},
  year={1995},
  publisher={Springer}
}

@article{schekochihin2022mhd,
  title={MHD turbulence: a biased review},
  author={Schekochihin, Alexander A},
  journal={Journal of Plasma Physics},
  volume={88},
  number={5},
  pages={155880501},
  year={2022},
  publisher={Cambridge University Press}
}

@book{gary1993theory,
  title={Theory of space plasma microinstabilities},
  author={Gary, S Peter},
  number={7},
  year={1993},
  publisher={Cambridge university press}
}

@article{klein2021inferred,
  title={Inferred linear stability of parker solar Probe observations using one-and two-component proton distributions},
  author={Klein, KG and Verniero, JL and Alterman, B and Bale, S and Case, A and Kasper, JC and Korreck, K and Larson, D and Lichko, E and Livi, R and others},
  journal={The Astrophysical Journal},
  volume={909},
  number={1},
  pages={7},
  year={2021},
  publisher={IOP Publishing}
}

@article{howes2010interpretation,
  title={On the interpretation of magnetic helicity signatures in the dissipation range of solar wind turbulence},
  author={Howes, Gregory G and Quataert, Eliot},
  journal={The Astrophysical Journal Letters},
  volume={709},
  number={1},
  pages={L49--L52},
  year={2010},
  publisher={The American Astronomical Society}
}

@article{squire2022high,
  title={High-frequency heating of the solar wind triggered by low-frequency turbulence},
  author={Squire, Jonathan and Meyrand, Romain and Kunz, Matthew W and Arzamasskiy, Lev and Schekochihin, Alexander A and Quataert, Eliot},
  journal={Nature Astronomy},
  volume={6},
  number={6},
  pages={715--723},
  year={2022},
  publisher={Nature Publishing Group UK London}
}

@article{meyrand2021violation,
  title={On the violation of the zeroth law of turbulence in space plasmas},
  author={Meyrand, Romain and Squire, Jonathan and Schekochihin, Alexander A and Dorland, William},
  journal={Journal of Plasma Physics},
  volume={87},
  number={3},
  pages={535870301},
  year={2021},
  publisher={Cambridge University Press}
}

@article{squire2023electron,
  title={Electron--ion heating partition in imbalanced solar-wind turbulence},
  author={Squire, Jonathan and Meyrand, Romain and Kunz, Matthew W},
  journal={The Astrophysical Journal Letters},
  volume={957},
  number={2},
  pages={L30},
  year={2023},
  publisher={The American Astronomical Society}
}

@article{chen2013nature,
  title={Nature of subproton scale turbulence in the solar wind},
  author={Chen, CHK and Boldyrev, S and Xia, Q and Perez, JC},
  journal={Physical review letters},
  volume={110},
  number={22},
  pages={225002},
  year={2013},
  publisher={APS}
}

@article{cerri2019kinetic,
  title={Kinetic plasma turbulence: recent insights and open questions from 3D3V simulations},
  author={Cerri, Silvio Sergio and Groselj, Daniel and Franci, Luca},
  journal={Frontiers in Astronomy and Space Sciences},
  volume={6},
  pages={64},
  year={2019},
  publisher={Frontiers}
}

@article{hellinger2006solar,
  title={Solar wind proton temperature anisotropy: Linear theory and WIND/SWE observations},
  author={Hellinger, Petr and Tr{\'a}vn{\'\i}{\v{c}}ek, Pavel and Kasper, Justin C and Lazarus, Alan J},
  journal={Geophysical research letters},
  volume={33},
  number={9},
  year={2006},
  publisher={Wiley Online Library}
}

@article{higdon1984density,
  title={Density fluctuations in the interstellar medium: Evidence for anisotropic magnetogasdynamic turbulence. I-Model and astrophysical sites},
  author={Higdon, JC},
  journal={Astrophysical Journal, Part 1 (ISSN 0004-637X), vol. 285, Oct. 1, 1984, p. 109-123.},
  volume={285},
  pages={109--123},
  year={1984}
}

@article{leamon1998observational,
  title={Observational constraints on the dynamics of the interplanetary magnetic field dissipation range},
  author={Leamon, Robert J and Smith, Charles W and Ness, Norman F and Matthaeus, William H and Wong, Hung K},
  journal={Journal of Geophysical Research: Space Physics},
  volume={103},
  number={A3},
  pages={4775--4787},
  year={1998},
  publisher={Wiley Online Library}
}

@article{sahraoui2009evidence,
  title={Evidence of a cascade and dissipation of solar-wind turbulence at the electron gyroscale},
  author={Sahraoui, Fouad and Goldstein, ML and Robert, Patrick and Khotyaintsev, Yu V},
  journal={Physical review letters},
  volume={102},
  number={23},
  pages={231102},
  year={2009},
  publisher={APS}
}

@article{kiyani2013enhanced,
  title={Enhanced magnetic compressibility and isotropic scale invariance at sub-ion Larmor scales in solar wind turbulence},
  author={Kiyani, Khurom H and Chapman, Sandra C and Sahraoui, Fouad and Hnat, Bogdan and Fauvarque, Olivier and Khotyaintsev, Yu V},
  journal={The Astrophysical Journal},
  volume={763},
  number={1},
  pages={10},
  year={2013},
  publisher={The American Astronomical Society}
}

@article{salem2012identification,
  title={Identification of kinetic Alfv{\'e}n wave turbulence in the solar wind},
  author={Salem, Chadi S and Howes, GG and Sundkvist, D and Bale, SD and Chaston, CC and Chen, CHK and Mozer, FS},
  journal={The Astrophysical Journal Letters},
  volume={745},
  number={1},
  pages={L9},
  year={2012},
  publisher={The American Astronomical Society}
}

@article{schekochihin2009astrophysical,
  title={Astrophysical gyrokinetics: kinetic and fluid turbulent cascades in magnetized weakly collisional plasmas},
  author={Schekochihin, AA and Cowley, Steven Charles and Dorland, W and Hammett, GW and Howes, Gregory G and Quataert, E and Tatsuno, T},
  journal={The Astrophysical Journal Supplement Series},
  volume={182},
  number={1},
  pages={310--377},
  year={2009},
  publisher={The American Astronomical Society}
}

@article{howes2008model,
  title={A model of turbulence in magnetized plasmas: Implications for the dissipation range in the solar wind},
  author={Howes, Gregory G and Cowley, Steven C and Dorland, William and Hammett, Gregory W and Quataert, Eliot and Schekochihin, Alexander A},
  journal={Journal of Geophysical Research: Space Physics},
  volume={113},
  number={A5},
  year={2008},
  publisher={Wiley Online Library}
}

@article{gary2004kinetic,
  title={Kinetic Alfv{\'e}n waves: Linear theory and a particle-in-cell simulation},
  author={Gary, S Peter and Nishimura, Kazumi},
  journal={Journal of Geophysical Research: Space Physics},
  volume={109},
  number={A2},
  year={2004},
  publisher={Wiley Online Library}
}

@article{vasconez2015kinetic,
  title={Kinetic Alfv{\'e}n wave generation by large-scale phase mixing},
  author={V{\'a}sconez, CL and Pucci, F and Valentini, Francesco and Servidio, Sergio and Matthaeus, WH and Malara, Francesco},
  journal={The Astrophysical Journal},
  volume={815},
  number={1},
  pages={7},
  year={2015},
  publisher={The American Astronomical Society}
}

@article{cerri2016subproton,
  title={Subproton-scale cascades in solar wind turbulence: driven hybrid-kinetic simulations},
  author={Cerri, SILVIO SERGIO and Califano, Francesco and Jenko, F and Told, D and Rincon, F},
  journal={The Astrophysical Journal Letters},
  volume={822},
  number={1},
  pages={L12},
  year={2016},
  publisher={The American Astronomical Society}
}

@article{grovselj2019kinetic,
  title={Kinetic turbulence in astrophysical plasmas: waves and/or structures?},
  author={Gro{\v{s}}elj, Daniel and Chen, Christopher HK and Mallet, Alfred and Samtaney, Ravi and Schneider, Kai and Jenko, Frank},
  journal={Physical Review X},
  volume={9},
  number={3},
  pages={031037},
  year={2019},
  publisher={APS}
}

@article{grovselj2018fully,
  title={Fully kinetic simulation of 3D kinetic Alfv{\'e}n turbulence},
  author={Gro{\v{s}}elj, Daniel and Mallet, Alfred and Loureiro, Nuno F and Jenko, Frank},
  journal={Physical review letters},
  volume={120},
  number={10},
  pages={105101},
  year={2018},
  publisher={APS}
}

@article{bowen2020ion,
  title={Ion-scale electromagnetic waves in the inner heliosphere},
  author={Bowen, Trevor A and Mallet, Alfred and Huang, Jia and Klein, Kristopher G and Malaspina, David M and Stevens, Michael and Bale, Stuart D and Bonnell, John W and Case, Anthony W and Chandran, Benjamin DG and others},
  journal={The Astrophysical Journal Supplement Series},
  volume={246},
  number={2},
  pages={66},
  year={2020},
  publisher={The American Astronomical Society}
}

@article{zhou2023electron,
  title={Electron heating in kinetic-Alfv{\'e}n-wave turbulence},
  author={Zhou, Muni and Liu, Zhuo and Loureiro, Nuno F},
  journal={Proceedings of the National Academy of Sciences},
  volume={120},
  number={23},
  pages={e2220927120},
  year={2023},
  publisher={National Academy of Sciences}
}

@article{liu2023radial,
  title={The radial distribution of ion-scale waves in the inner heliosphere},
  author={Liu, Wen and Zhao, Jinsong and Wang, Tieyan and Dong, Xiangcheng and Kasper, Justin C and Bale, Stuart D and Shi, Chen and Wu, Dejin},
  journal={The Astrophysical Journal},
  volume={951},
  number={1},
  pages={69},
  year={2023},
  publisher={The American Astronomical Society}
}

@article{zhao2021mhd,
  title={MHD and ion kinetic waves in field-aligned flows observed by Parker solar probe},
  author={Zhao, L-L and Zank, GP and He, JS and Telloni, D and Adhikari, L and Nakanotani, M and Kasper, JC and Bale, SD},
  journal={The Astrophysical Journal},
  volume={922},
  number={2},
  pages={188},
  year={2021},
  publisher={The American Astronomical Society}
}

@article{leamon1999dissipation,
  title={Dissipation range dynamics: Kinetic Alfv{\'e}n waves and the importance of $\beta$ e},
  author={Leamon, Robert J and Smith, Charles W and Ness, Norman F and Wong, Hung K},
  journal={Journal of Geophysical Research: Space Physics},
  volume={104},
  number={A10},
  pages={22331--22344},
  year={1999},
  publisher={Wiley Online Library}
}

@article{quataert1998particle,
  title={Particle heating by Alfv{\'e}nic turbulence in hot accretion flows},
  author={Quataert, Eliot},
  journal={The Astrophysical Journal},
  volume={500},
  number={2},
  pages={978--991},
  year={1998}
}

@phdthesis{barnes1966collisionless,
  title={Collisionless damping of hydromagnetic waves},
  author={Barnes, Aaron},
  year={1966},
  school={The University of Chicago}
}

@article{chandran2010perpendicular,
  title={Perpendicular ion heating by low-frequency Alfv{\'e}n-wave turbulence in the solar wind},
  author={Chandran, Benjamin DG and Li, Bo and Rogers, Barrett N and Quataert, Eliot and Germaschewski, Kai},
  journal={The Astrophysical Journal},
  volume={720},
  number={1},
  pages={503--515},
  year={2010},
  publisher={The American Astronomical Society}
}

@article{hollweg1999kinetic,
  title={Kinetic Alfv{\'e}n wave revisited},
  author={Hollweg, Joseph V},
  journal={Journal of Geophysical Research: Space Physics},
  volume={104},
  number={A7},
  pages={14811--14819},
  year={1999},
  publisher={Wiley Online Library}
}

@article{hollweg2002generation,
  title={Generation of the fast solar wind: A review with emphasis on the resonant cyclotron interaction},
  author={Hollweg, Joseph V and Isenberg, Philip A},
  journal={Journal of Geophysical Research: Space Physics},
  volume={107},
  number={A7},
  pages={SSH--12},
  year={2002},
  publisher={Wiley Online Library}
}

@article{isenberg2011kinetic,
  title={A kinetic model of solar wind generation by oblique ion-cyclotron waves},
  author={Isenberg, Philip A and Vasquez, Bernard J},
  journal={The Astrophysical Journal},
  volume={731},
  number={2},
  pages={88},
  year={2011},
  publisher={The American Astronomical Society}
}

@article{isenberg2019perpendicular,
  title={Perpendicular ion heating by cyclotron resonant dissipation of turbulently generated kinetic Alfv{\'e}n waves in the solar wind},
  author={Isenberg, Philip A and Vasquez, Bernard J},
  journal={The Astrophysical Journal},
  volume={887},
  number={1},
  pages={63},
  year={2019},
  publisher={The American Astronomical Society}
}

@article{hoppock2018stochastic,
  title={Stochastic proton heating by kinetic-Alfv{\'e}n-wave turbulence in moderately high-plasmas},
  author={Hoppock, Ian W and Chandran, Benjamin DG and Klein, Kristopher G and Mallet, Alfred and Verscharen, Daniel},
  journal={Journal of plasma physics},
  volume={84},
  number={6},
  pages={905840615},
  year={2018},
  publisher={Cambridge University Press}
}

@article{osman2012kinetic,
  title={Kinetic signatures and intermittent turbulence in the solar wind plasma},
  author={Osman, KT and Matthaeus, WH and Hnat, B and Chapman, Sandra C},
  journal={Physical Review Letters},
  volume={108},
  number={26},
  pages={261103},
  year={2012},
  publisher={APS}
}

@article{dmitruk2004test,
  title={Test particle energization by current sheets and nonuniform fields in magnetohydrodynamic turbulence},
  author={Dmitruk, Pablo and Matthaeus, WH and Seenu, N},
  journal={The Astrophysical Journal},
  volume={617},
  number={1},
  pages={667--679},
  year={2004}
}

@article{mallet2026perpendicular,
  title={Perpendicular ion heating in turbulence and reconnection: magnetic moment breaking by coherent fluctuations},
  author={Mallet, Alfred and Klein, Kristopher and Chandran, Benjamin Divakar Giles and Ervin, Tamar and Bowen, Trevor A},
  journal={Journal of Plasma Physics},
  volume={92},
  number={1},
  pages={E13},
  year={2026},
  publisher={Cambridge University Press}
}

@article{woodham2019parallel,
  title={Parallel-propagating fluctuations at proton-kinetic scales in the solar wind are dominated by kinetic instabilities},
  author={Woodham, Lloyd D and Wicks, Robert T and Verscharen, Daniel and Owen, Christopher J and Maruca, Bennett A and Alterman, Benjamin L},
  journal={The Astrophysical Journal Letters},
  volume={884},
  number={2},
  pages={L53},
  year={2019},
  publisher={The American Astronomical Society}
}

@article{liu2025effects,
  title={Effects of Ion Temperature Anisotropy on Ion-scale Waves’ Generation in the Near-Sun Solar Wind: Parker Solar Probe Observations},
  author={Liu, Wen and Jia, Huan-Yu and Liu, Si-Ming},
  journal={The Astrophysical Journal Letters},
  volume={983},
  number={2},
  pages={L35},
  year={2025},
  publisher={The American Astronomical Society}
}

@article{niranjana2026free,
  title={Free Energy Sources of Ion-scale Waves Observed by Parker Solar Probe},
  author={Niranjana and Klein, Kristopher G and Martinovi{\'c}, Mihailo M and Bowen, Trevor A and Larson, Davin E and Livi, Roberto and Rahmati, Ali and Whittlesey, Phyllis L and Stevens, Michael L},
  journal={The Astrophysical Journal Supplement Series},
  volume={284},
  number={1},
  pages={4},
  year={2026},
  publisher={The American Astronomical Society}
}

@article{schekochihin2019constraints,
  title={Constraints on ion versus electron heating by plasma turbulence at low beta},
  author={Schekochihin, AA and Kawazura, Y and Barnes, MA},
  journal={Journal of Plasma Physics},
  volume={85},
  number={3},
  pages={905850303},
  year={2019},
  publisher={Cambridge University Press}
}

@article{zhang2025extreme,
  title={Extreme heating of minor ions in imbalanced solar-wind turbulence},
  author={Zhang, Michael F and Kunz, Matthew W and Squire, Jonathan and Klein, Kristopher G},
  journal={The Astrophysical Journal},
  volume={979},
  number={2},
  pages={121},
  year={2025},
  publisher={The American Astronomical Society}
}

@article{arzamasskiy2019hybrid,
  title={Hybrid-kinetic simulations of ion heating in Alfv{\'e}nic turbulence},
  author={Arzamasskiy, Lev and Kunz, Matthew W and Chandran, Benjamin DG and Quataert, Eliot},
  journal={The Astrophysical Journal},
  volume={879},
  number={1},
  pages={53},
  year={2019},
  publisher={The American Astronomical Society}
}

@article{greco2009statistical,
  title={Statistical analysis of discontinuities in solar wind ACE data and comparison with intermittent MHD turbulence},
  author={Greco, Antonella and Matthaeus, WH and Servidio, Sergio and Chuychai, P and Dmitruk, P},
  journal={The Astrophysical Journal},
  volume={691},
  number={2},
  pages={L111--L114},
  year={2009},
  publisher={The American Astronomical Society}
}

@article{iroshnikov1963turbulence,
  title={Turbulence of a conducting fluid in a strong magnetic field},
  author={Iroshnikov, PS},
  journal={Astronomicheskii Zhurnal},
  volume={40},
  pages={742},
  year={1963}
}

@article{kraichnan1965inertial,
  title={Inertial-range spectrum of hydromagnetic turbulence},
  author={Kraichnan, Robert H},
  journal={Physics of Fluids},
  volume={8},
  number={7},
  pages={1385},
  year={1965}
}

@article{gonzalez2024local,
  title={Local proton heating at magnetic discontinuities in Alfv{\'e}nic and non-Alfv{\'e}nic solar wind},
  author={Gonz{\'a}lez, CA and Verniero, JL and Bandyopadhyay, R and Tenerani, A},
  journal={The Astrophysical Journal},
  volume={963},
  number={2},
  pages={148},
  year={2024},
  publisher={The American Astronomical Society}
}

@article{kennel1966velocity,
  title={Velocity space diffusion from weak plasma turbulence in a magnetic field},
  author={Kennel, CF and Engelmann, F},
  journal={Physics of Fluids},
  volume={9},
  number={12},
  pages={2377},
  year={1966}
}

@book{stix1992waves,
  title={Waves in plasmas},
  author={Stix, Thomas H},
  year={1992},
  publisher={Springer Science \& Business Media}
}

@article{buti2000hybrid,
  title={Hybrid simulations of collapse of Alfv{\'e}nic wave packets},
  author={Buti, B and Velli, Marco and Liewer, PC and Goldstein, Bruce E and Hada, T},
  journal={Physics of Plasmas},
  volume={7},
  number={10},
  pages={3998--4003},
  year={2000},
  publisher={American Institute of Physics}
}

@article{machida_simulation_1987,
	title = {Simulation of amplitude-modulated circularly polarized {Alfvén} waves for beta less than one},
	volume = {92},
	issn = {2156-2202},
	language = {en},
	number = {A7},
	urldate = {2025-05-27},
	journal = {Journal of Geophysical Research: Space Physics},
	author = {Machida, S. and Spangler, S. R. and Goertz, C. K.},
	year = {1987},
	note = {\_eprint: https://agupubs.onlinelibrary.wiley.com/doi/pdf/10.1029/JA092iA07p07413},
	pages = {7413--7422}
}

@article{markovskii2023magnetic,
  title={Magnetic Helicity Associated with the Proton Temperature Anisotropy Instabilities in the Presence of an Imbalanced Solar Wind Turbulence},
  author={Markovskii, SA and Vasquez, Bernard J},
  journal={The Astrophysical Journal},
  volume={952},
  number={1},
  pages={16},
  year={2023},
  publisher={The American Astronomical Society}
}

@article{elmegreen2004interstellar,
  title={Interstellar turbulence I: observations and processes},
  author={Elmegreen, Bruce G and Scalo, John},
  journal={Annu. Rev. Astron. Astrophys.},
  volume={42},
  number={1},
  pages={211--273},
  year={2004},
  publisher={Annual Reviews}
}

@incollection{schekochihin2007turbulence,
  title={Turbulence and magnetic fields in astrophysical plasmas},
  author={Schekochihin, Alexander A and Cowley, Steven C},
  booktitle={Magnetohydrodynamics: Historical Evolution and Trends},
  pages={85--115},
  year={2007},
  publisher={Springer}
}

@article{jian2014electromagnetic,
  title={Electromagnetic waves near the proton cyclotron frequency: STEREO observations},
  author={Jian, LK and Wei, HY and Russell, CT and Luhmann, JG and Klecker, B and Omidi, N and Isenberg, PA and Goldstein, ML and Figueroa-Vi{\~n}as, A and Blanco-Cano, X},
  journal={The Astrophysical Journal},
  volume={786},
  number={2},
  pages={123},
  year={2014},
  publisher={The American Astronomical Society}
}

@article{jian2009ion,
  title={Ion cyclotron waves in the solar wind observed by STEREO near 1 AU},
  author={Jian, Lan K and Russell, Christopher T and Luhmann, Janet G and Strangeway, Robert J and Leisner, Jared S and Galvin, Antoinette B},
  journal={The Astrophysical Journal},
  volume={701},
  number={2},
  pages={L105--L109},
  year={2009},
  publisher={The American Astronomical Society}
}

@article{bianco2026evolution,
  title={Evolution of an Alfv{\'e}n Wave--driven Proton Beam in the Expanding Solar Wind},
  author={Bianco, JS and Tenerani, A and Gonzalez, C and Matteini, L and Klein, KG},
  journal={The Astrophysical Journal},
  volume={998},
  number={2},
  pages={194},
  year={2026},
  publisher={The American Astronomical Society}
}

@article{bowen2024mediation,
  title={Mediation of collisionless turbulent dissipation through cyclotron resonance},
  author={Bowen, Trevor A and Bale, Stuart D and Chandran, Benjamin DG and Chasapis, Alexandros and Chen, Christopher HK and Dudok de Wit, Thierry and Mallet, Alfred and Meyrand, Romain and Squire, Jonathan},
  journal={Nature astronomy},
  volume={8},
  number={4},
  pages={482--490},
  year={2024},
  publisher={Nature Publishing Group UK London}
}

@article{johnston2026quasi,
  title={Quasi-linear theory of perpendicular ion heating by critically balanced turbulence},
  author={Johnston, Zade and Squire, Jonathan},
  journal={Journal of Plasma Physics},
  volume={92},
  number={3},
  pages={E73},
  year={2026},
  publisher={Cambridge University Press}
}

@article{mallet2019interplay,
  title={Interplay between intermittency and dissipation in collisionless plasma turbulence},
  author={Mallet, Alfred and Klein, Kristopher G and Chandran, Benjamin DG and Gro{\v{s}}elj, Daniel and Hoppock, Ian W and Bowen, Trevor A and Salem, Chadi S and Bale, Stuart D},
  journal={Journal of Plasma Physics},
  volume={85},
  number={3},
  pages={175850302},
  year={2019},
  publisher={Cambridge University Press}
}

@article{chandran2026second,
  title={The Second Quadratic Invariant of Astrophysical Gyrokinetics},
  author={Chandran, Benjamin DG and Mallet, Alfred and Meyrand, Romain},
  year={2026}
}

@article{velli1989turbulent,
  title={Turbulent cascade of incompressible unidirectional Alfv{\'e}n waves in the interplanetary medium},
  author={Velli, Marco and Grappin, Roland and Mangeney, Andr{\'e}},
  journal={Physical Review Letters},
  volume={63},
  number={17},
  pages={1807},
  year={1989},
  publisher={APS}
}

@article{chandran2009alfven,
  title={Alfv{\'e}n wave reflection and turbulent heating in the solar wind from 1 solar radius to 1 AU: an analytical treatment},
  author={Chandran, Benjamin DG and Hollweg, Joseph V},
  journal={The Astrophysical Journal},
  volume={707},
  number={2},
  pages={1659--1667},
  year={2009},
  publisher={The American Astronomical Society}
}

@article{meyrand2025reflection,
  title={Reflection-driven turbulence in the super-Alfv{\'e}nic solar wind},
  author={Meyrand, Romain and Squire, Jonathan and Mallet, Alfred and Chandran, Benjamin DG},
  journal={Journal of plasma physics},
  volume={91},
  number={1},
  pages={E29},
  year={2025},
  publisher={Cambridge University Press}
}

@article{gonzalez2019turbulent,
  title={Turbulent electromagnetic fields at sub-proton scales: Two-fluid and full-kinetic plasma simulations},
  author={Gonz{\'a}lez, CA and Parashar, TN and Gomez, D and Matthaeus, WH and Dmitruk, P},
  journal={Physics of Plasmas},
  volume={26},
  number={1},
  year={2019},
  publisher={AIP Publishing}
}

@article{goldreich1995toward,
  title={Toward a theory of interstellar turbulence. 2: Strong alfvenic turbulence},
  author={Goldreich, P and Sridhar, S},
  journal={Astrophysical Journal, Part 1 (ISSN 0004-637X), vol. 438, no. 2, p. 763-775},
  volume={438},
  pages={763--775},
  year={1995}
}

@article{galtier2000weak,
  title={A weak turbulence theory for incompressible magnetohydrodynamics},
  author={Galtier, S{\'e}bastien and Nazarenko, SV and Newell, Alan C and Pouquet, Annick},
  journal={Journal of plasma physics},
  volume={63},
  number={5},
  pages={447--488},
  year={2000},
  publisher={Cambridge University Press}
}

@article{dmitruk2006structure,
  title={Structure of the electromagnetic field in three-dimensional Hall magnetohydrodynamic turbulence},
  author={Dmitruk, Pablo and Matthaeus, WH},
  journal={Physics of Plasmas},
  volume={13},
  number={4},
  year={2006},
  publisher={AIP Publishing}
}

@article{zhao2026ion,
  title={Ion-scale Waves Regulated by Plasma Beta and Cross Helicity: Observational Evidence for the Helicity Barrier in Imbalanced Solar Wind Turbulence},
  author={Zhao, GQ and Meyrand, R and Feng, HQ and Yang, HF and Xiang, L and Wang, Z},
  journal={The Astrophysical Journal Letters},
  volume={1008},
  number={2},
  pages={L34},
  year={2026},
  publisher={The American Astronomical Society}
}

@article{Qian_2026,
year = {2026},
month = {sep},
publisher = {The American Astronomical Society},
volume = {1008},
number = {2},
pages = {244},
author = {Qian, Hai Yang Harry and Bowen, Trevor A. and Gonzalez, Carlos A. and Sioulas, Nikos and Mallet, Alfred and Klein, Kristopher G. and Verscharen, Daniel and Bale, Stuart D.},
title = {Physics of Circularly Polarized Ion-scale Waves in Hybrid Simulations of Alfvénic Fluctuations},
journal = {The Astrophysical Journal}
}

@article{mcintyre2025,
  title={Evidence for the helicity barrier from measurements of the turbulence transition range in the solar wind},
  author={McIntyre, JR and Chen, Christopher HK and Squire, Jonathan and Meyrand, Romain and Simon, PA},
  journal={Physical Review X},
  volume={15},
  number={3},
  pages={031008},
  year={2025},
  publisher={APS}
}

@ARTICLE{2025PhRvL.135y5201B,
       author = {{Bowen}, Trevor A. and {Ervin}, Tamar and {Mallet}, Alfred and {Chandran}, Benjamin D.~G. and {Sioulas}, Nikos and {Isenberg}, Philip A. and {Bale}, Stuart D. and {Squire}, Jonathan and {Klein}, Kristopher G. and {Pezzi}, Oreste},
        title = "{Stochastic Heating in the Sub-Alfv{\'e}nic Solar Wind}",
      journal = {\prl},
         year = 2025,
        month = dec,
       volume = {135},
       number = {25},
          eid = {255201},
        pages = {255201},
 primaryClass = {astro-ph.SR},
       adsurl = {https://ui.adsabs.harvard.edu/abs/2025PhRvL.135y5201B}
}

@article{chandran2010,
  title={Resonant interactions between protons and oblique Alfv{\'e}n/ion-cyclotron waves in the solar corona and solar flares},
  author={Chandran, Benjamin DG and Pongkitiwanichakul, Peera and Isenberg, Philip A and Lee, Martin A and Markovskii, Sergei A and Hollweg, Joseph V and Vasquez, Bernard J},
  journal={The Astrophysical Journal},
  volume={722},
  number={1},
  pages={710--720},
  year={2010},
  publisher={The American Astronomical Society}
}

@article{chandran2019reflection,
  title={Reflection-driven magnetohydrodynamic turbulence in the solar atmosphere and solar wind},
  author={Chandran, Benjamin DG and Perez, Jean C},
  journal={Journal of Plasma Physics},
  volume={85},
  number={4},
  pages={905850409},
  year={2019},
  publisher={Cambridge University Press}
}

@article{sahraoui2010three,
  title={Three dimensional anisotropic k spectra of turbulence at subproton scales in the solar wind},
  author={Sahraoui, Fouad and Goldstein, Melvyn L and Belmont, G{\'e}rard and Canu, Patrick and Rezeau, Laurence},
  journal={Physical review letters},
  volume={105},
  number={13},
  pages={131101},
  year={2010},
  publisher={APS}
}

@article{bowen2020constraining,
  title={Constraining ion-scale heating and spectral energy transfer in observations of plasma turbulence},
  author={Bowen, Trevor A and Mallet, Alfred and Bale, Stuart D and Bonnell, JW and Case, Anthony W and Chandran, Benjamin DG and Chasapis, Alexandros and Chen, Christopher HK and Duan, Die and Dudok de Wit, Thierry and others},
  journal={Physical review letters},
  volume={125},
  number={2},
  pages={025102},
  year={2020},
  publisher={APS}
}

@article{podesta2011magnetic,
  title={Magnetic helicity spectrum of solar wind fluctuations as a function of the angle with respect to the local mean magnetic field},
  author={Podesta, John J and Gary, S Peter},
  journal={The Astrophysical Journal},
  volume={734},
  number={1},
  pages={15},
  year={2011},
  publisher={The American Astronomical Society}
}

@article{he2011possible,
  title={Possible evidence of Alfv{\'e}n-cyclotron waves in the angle distribution of magnetic helicity of solar wind turbulence},
  author={He, Jiansen and Marsch, Eckart and Tu, Chuanyi and Yao, Shuo and Tian, Hui},
  journal={The Astrophysical Journal},
  volume={731},
  number={2},
  pages={85},
  year={2011},
  publisher={The American Astronomical Society}
}

@article{huang2020kinetic,
  title={Kinetic scale slow solar wind turbulence in the inner heliosphere: coexistence of kinetic Alfv{\'e}n waves and Alfv{\'e}n ion cyclotron waves},
  author={Huang, SY and Zhang, J and Sahraoui, F and He, JS and Yuan, ZG and Andr{\'e}s, Nahuel and Hadid, LZ and Deng, XH and Jiang, K and Yu, L and others},
  journal={The Astrophysical Journal Letters},
  volume={897},
  number={1},
  pages={L3},
  year={2020},
  publisher={The American Astronomical Society}
}

@article{told2016linear,
  title={A linear dispersion relation for the hybrid kinetic-ion/fluid-electron model of plasma physics},
  author={Told, Daniel and Cookmeyer, J and Astfalk, Patrick and Jenko, Frank},
  journal={New Journal of Physics},
  volume={18},
  number={7},
  pages={075001},
  year={2016},
  publisher={IOP Publishing}
}

@article{kobayashi2017three,
  title={Three-dimensional simulations and spacecraft observations of sub-ion scale turbulence in the solar wind: influence of Landau damping},
  author={Kobayashi, Sumire and Sahraoui, Fouad and Passot, T and Laveder, D and Sulem, PL and Huang, SY and Henri, Pierre and Smets, R},
  journal={The Astrophysical Journal},
  volume={839},
  number={2},
  pages={122},
  year={2017},
  publisher={The American Astronomical Society}
}

@article{bowen2025stochastic,
  title={Stochastic heating in the sub-Alfv{\'e}nic solar wind},
  author={Bowen, Trevor A and Ervin, Tamar and Mallet, Alfred and Chandran, Benjamin DG and Sioulas, Nikos and Isenberg, Philip A and Bale, Stuart D and Squire, Jonathan and Klein, Kristopher G and Pezzi, Oreste},
  journal={Physical review letters},
  volume={135},
  number={25},
  pages={255201},
  year={2025},
  publisher={APS}
}

@article{ervin2026direct,
  title={Direct Measurement of Diffusion Coefficients: Evidence for Diffusive Stochastic Heating in Collisionless Plasmas},
  author={Ervin, Tamar and Bowen, Trevor A and Mallet, Alfred and Isenberg, Philip A and Klein, Kristopher G and Bale, Stuart D and Chandran, Benjamin DG and Livi, Roberto and Rahmati, Ali and Larson, Davin E},
  journal={Physical Review Letters},
  volume={137},
  number={6},
  pages={065201},
  year={2026},
  publisher={APS}
}

@article{klein2016measuring,
  title={Measuring collisionless damping in heliospheric plasmas using field--particle correlations},
  author={Klein, Kristopher G and Howes, Gregory G},
  journal={The Astrophysical Journal Letters},
  volume={826},
  number={2},
  pages={L30},
  year={2016},
  publisher={The American Astronomical Society}
}

@article{gonzalez2017test,
  title={Test particle energization and the anisotropic effects of dynamical MHD turbulence},
  author={Gonz{\'a}lez, CA and Dmitruk, P and Mininni, Pablo Daniel and Matthaeus, WH},
  journal={The Astrophysical Journal},
  volume={850},
  number={1},
  pages={19},
  year={2017},
  publisher={The American Astronomical Society}
}
\bibliographystyle{aasjournalv7}

\end{document}